\documentclass[fleqn,usenatbib]{mnras}
\usepackage{newtxtext,newtxmath}
\usepackage[T1]{fontenc}

\DeclareRobustCommand{\VAN}[3]{#2}
\let\VANthebibliography\thebibliography
\def\thebibliography{\DeclareRobustCommand{\VAN}[3]{##3}\VANthebibliography}

\usepackage{graphicx}	
\usepackage{amsmath}	

\newcommand*  {\diff}       {\mathop{}\!\mathrm{d}}

\newcommand*  {\Sect}[1]    {Section~\ref{#1}}

\renewcommand*{\vec}[1]     {\boldsymbol{#1}}

\newcommand*  {\p}          {\partial}

\newcommand*  {\pdiff}[2]   {\frac{\p{#1}}{\p{#2}}}
\newcommand*  {\tdiff}[2]   {\frac{\diff{#1}}{\diff{#2}}}

\newcommand*  {\sub}[2]     {{#1}_{\mathrm{#2}}}
\newcommand*  {\au}         {{\mathrm{au}}}
\newcommand*  {\kms}        {{\mathrm{km\,s^{-1}}}}

\title{Capture of interstellar objects II: by the Solar system}

\author[Dehnen, Hands, Sch{\"o}nrich]{%
Walter~Dehnen,$^{\!\!\!1,2,3}$
Thomas~O.~Hands,$^{\!\!4}$ and
Ralph Sch{\"o}nrich$^{5}$\smallskip
\\
$^1$ Astronomisches Recheninstitut, Zentrum f{\"u}r Astronomie der Universit{\"a}t Heidelberg, M{\"o}nchhofstra\ss{}e. 12-14, 69120, Heidelberg, Germany\\
$^2$ Universit{\"a}ts-Sternwarte M{\"u}nchen, Scheinerstra\ss{}e 1, 81679, M{\"u}nchen, Germany\\
$^3$ School for Physics and Astronomy, University of Leicester, University Road, LE1 7RH, UK\\
$^4$ Institut f{\"u}r Computergest{\"u}tzte Wissenschaften, Universit{\"a}t Z{\"u}rich, Winterthurerstrasse 190, CH-8057 Z{\"u}rich, Switzerland\\
$^5$ Mullard Space Science Laboratory, University College London, Holmbury St.~Mary, Dorking, Surrey, RH5 6NT, UK
}

\date{Accepted XXX. Received YYY; in original form ZZZ}

\pubyear{2021}

\begin{document}

\defcitealias{paper1}{paper~1}
\label{firstpage}
\pagerange{\pageref{firstpage}--\pageref{lastpage}}
\maketitle

\begin{abstract}
Capture of interstellar objects (ISOs) into the Solar system is dominated by ISOs with asymptotic incoming speeds $v_\infty<4\,$km\,s$^{-1}$. The capture rate is proportional to the ISO phase-space density in the Solar vicinity and does not vary along the Sun's Galactic orbit, i.e.\ is not enhanced during a passage through a cloud of ISOs (in contrast to previous suggestions). Most bound orbits crossing those of Jupiter and Saturn are fully mixed with unbound phase space, implying that they hold the same ISO phase-space density. Assuming an interstellar number density $\sub{n}{iso}\sim0.1\,$au$^{-3}$, we estimate that in 1000 years the planets capture $\sim2$ ISOs (while $\sim17$ fall into the Sun), resulting in a population of $\sim8$ captured ISOs within 5\,au of the Sun at any time, less than the number of visiting ISOs passing through the same volume on hyperbolic orbits. In terms of phase-space volume, capture onto and ejection from the Solar system are equal, such that on average ISOs will not remain captive at $a\lesssim2000\,$au for extensive periods.
\end{abstract}

\begin{keywords}
    celestial mechanics –– comets: general –– comets: individual: 2I/Borisov –– minor planets, asteroids: general –– minor planets, asteroids: individual: 1I/‘Oumuamua –– Oort Cloud.
\end{keywords}    

\section{Introduction}
In 2017, the first unquestionably interstellar object (ISO) passing through the Solar system was observed: 1I/2017 U1, later named 'Oumuamua \citep[][]{Meech2017,ISSI2019}, with a light curve that implies an extreme axis ratio between its longest and shortest axis (at the moment of its discovery one month after perihelion). Due to the lack of a detectable coma, 'Oumuamua was originally thought to be an asteroid of $\sim100\,$m size, but its orbital anomalies and highly elongated shape are most naturally explained if it is a $\sim40\,$m sized fragment of nitrogen ice \citep{Jackson21}, which is naturally created by an impact on the surface of a Pluto or Triton analogue \citep{Desch21}.  

Two years later, a second ISO was observed -- 2I/Borisov \citep{JewittLuu2019} -- this time exhibiting obvious cometary activity. The existence of such interstellar asteroids and comets has been hypothesised for decades \citep[e.g.,][]{Whipple1975, Sekania1976, Noerdlinger1977, ClubeNapier1984, ValtonenEtAl1992}. As messengers from other solar systems, they offer a wealth of information on their home systems beyond the possibility of directly probing ISOs: What ejects ISOs from their natal system, how are they incorporated into other planetary systems -- particularly our own? These two questions are inexorably linked: the physics of ejecting and capturing small bodies is very similar, and linked to a variety of phenomena in planetary systems.

The journey of ISOs begins in young planetary systems. Planetesimals left over from planet formation are readily ejected by Neptune and Jupiter analogues \citep[potentially being tidally stripped in the process][]{Raymond2018a}, and/or stripped from their parent stars in stellar fly-bys \citep[][]{HandsEtAl2019}. The volatile content of ISOs encodes where they formed in their parent system, e.g.\ 2I/Borisov must have been formed far from its parent star, like the Solar system comets. \cite{Raymond2018b} argued that outer-system planetesimals are more likely to become ISOs than inner-system planetesimals, because they are more abundant and more easily ejected due their lower binding energy (and suggested that 1I/'Oumuamua may not be of asteroidal composition, independently from the arguments for it being a made from nitrogen ice).
However, the details of this ratio will strongly depend on exo-planetary system architecture that determine the incidence of planetary chaos and ejections from the inner systems. This makes understanding capture, loss, and comparison to the (eventually) observed ratio an important diagnostic for the still unknown exo-planetary system architectures.

Like `Oumuamua, some passing ISO will visit the inner Solar system \citep{PortegiesZwart2018}, opening the door for direct observational investigations either telescopically \citep[e.g.][]{TrillingEtAl2018, GuzikEtAl2020} or by flyby, rendezvous, and sample-return spacecraft missions \citep[e.g.][]{SeligmanLaughlin2018, Hein2019, Hibberd2021}. This possibility coincides with the first observations of protoplanetary discs at exquisite resolution, e.g. by ALMA \citep{Alma2015}. In this context, the information from ISOs will be vital for filling in the missing pieces of the planet formation puzzle. The first two ISOs provide a tantalising hint to the wealth of information on chemistry and processes in exoplanetary systems: If we accept the explanations for 'Oumamua, it is actually the first fragment of a Pluto-like object. Remote observations of comet-like ISOs are in principle easier, as coma/tails allow for easy spectroscopy and thus chemical analysis \citep[see e.g.,][]{Fitzsimmons2019,McKay2020,Opitom2021}.

Another exciting possibility is to identify ISOs which have been captured by the Solar system long ago and are orbiting under our very noses. While there is currently no evidence that any known object in the Solar system is a captured ISO \citep{Morbi2020}, this may just be because we have not been looking in the correct places, but also because these exobodies are small with low albedo. In an accompanying paper~\citep[][hereafter paper~1]{paper1}, we have investigated, both analytically and numerically, the cross-section $\sigma$ for capturing an ISO by a planet-star binary.

The aim of this study is to apply the results of \citetalias{paper1} to calculate the ISO capture rate by the Solar system and predict the resident population of exobodies in the Solar system, including their semi-major axis distribution. Both the capture rate and the population of captive ISOs depend directly on the number density of these objects in interstellar space, which is not well known. From the meagre detection history prior to the discovery of 1I/'Oumuamua \cite{EngelhardtEtAl2017} estimated the number density in the Solar neighbourhood of $\sim1\,$km sized interstellar asteroids to be $n\sim0.02\,\mathrm{au}^{-3}$, while \cite{DoTuckerTonry2018} estimated  $n\sim0.2\,\mathrm{au}^{-3}$ for 'Oumuamua analogues (100\,m asteroids or 40\,m fragments of N$_2$ ice). However, in view of the lack of further discoveries, a more conservative assumption is $n\sim0.1\,\mathrm{au}^{-3}$ for such faint objects. For interstellar comets, the number density is likely $\sim100$ times smaller (since otherwise more would have been seen, \citeauthor{EngelhardtEtAl2017}), though \cite{SirajLoeb2021} recently estimated $n\sim0.009\,$au$^{-3}$ based on the discovery of 2I/Borisov. Consequently, we refrain from assuming a certain value, but keep $\sub{n}{iso}$ as parameter.

Previously, \citep{HandsDehnen2020} performed simulations of initially-unbound, massless test particles interacting with the Sun-Jupiter binary system: in a small minority of cases, low-velocity objects become bound, predicting a steady-state population of up to $~10^5$ `Oumuamua-style objects hiding in the outer Solar system. These numbers hint at the exciting prospect of studying an interstellar visitor in much greater detail. Here, we present a much more general study which considers both the ejection and capture of planetesimals by a planetary system with both analytical and numerical methods.

This paper is organised as follows. \Sect{sec:capture:sigma} briefly summarises some results of \citetalias{paper1}, which are applied in \Sect{sec:capture:rate} to calculate the rate at which ISOs are captured. In \Sect{sec:pop} we estimate the resulting steady-state population of captive ISOs in the Solar system, while \Sect{sec:conclude} summarises and concludes our paper.

\section{The capture cross section}
\label{sec:capture:sigma}
Here, we briefly summarise the relevant results of \citetalias{paper1}. Consider a planet-star binary with total mass $M=\sub{m}p+\sub{m}s$, mass ratio $q=\sub{m}p/\sub{m}s\ll1$, semi-major axis $\sub{a}p$, and eccentricity $\sub{e}p$. Then the cross-section for capturing an ISO with incoming asymptotic speed $v_\infty$ onto a bound orbit with semi-major axis $a\ge GM/\sub{v}a^2$ is 
\begin{align}
    \label{eq:sigma:theory}
	\sigma(v_\infty|\sub{v}a) = \pi\sub{a}p^2\,\frac{\sub{v}c^2}{v_\infty^2}\,f(X)\;
	Y\left(1,\frac{v_\infty^2}{\sub{v}c^2},\frac{\sub{v}a^2}{\sub{v}c^2}\right),
\end{align}
where $\sub{v}c^2=GM/\sub{a}p$, while
\begin{align}
    \label{eq:X}
    X \equiv 2|\Delta E|/q\sub{v}c^2
      = (v_\infty^2+\sub{v}a^2)/q\sub{v}c^2
\end{align}
is a dimensionless measure for the energy change $\Delta E$ required for capture. The function $f(X)$ in equation~\eqref{eq:sigma:theory} is determined empirically in \citetalias{paper1} (Fig.~5), and is approximated (to within $\sim10\%$) by 
\begin{align}
    \label{eq:f(X):fit}
    f(X) \approx \frac{8}{3X_0^2}\left[\sinh^{-1}(X_0/X)^{2/\alpha}\right]^{\alpha}
\end{align}
with $X_0\approx2.95$ and $\alpha\approx0.82$. At $X\lesssim1$, captures are dominated by wide encounters with the planet and $f(X)$ becomes proportional to $|\ln X|^\alpha$ at $X\ll1$. Conversely, for $X\gtrsim1$ only close planet encounters can provide sufficient $|\Delta E|$, and $f(X)$ approaches $\tfrac83X^{-2}$ at $X\gg1$.

Finally, the \emph{transfer function} $Y(1,x,z)$ in~\eqref{eq:sigma:theory} is given in equation~(26) of \citetalias{paper1} and plotted in its Fig.~2. For small values of its arguments, i.e.\ for $v_\infty\ll\sub{v}c$ and $\sub{v}a\ll\sub{v}c$, it deviates only weakly from its maximum $Y(1,0,0)=1$, but decays to zero at $v_\infty,\,\sub{v}a\gtrsim\sub{v}c$, corresponding to $X\gtrsim q^{-1}$, when even the closest possible encounters with the planet are insufficient. Planets of finite size cannot even deliver this, since capturing trajectories result in collisions instead, depending on the planet's radius $\sub{R}p$ as expressed by its Safronov number $\Theta=q\sub{a}p/\sub{R}p$. For the outer planets of the Solar system, however, $\Theta$ is large and this limitation largely irrelevant.

Very close encounters that just avoid collision may induce the tidal break-up of the ISO, depending on the Roche limit/relative densities of ISO and planet \citepalias[see also][Section 4.1]{paper1}, when some of its fragments can still become captured. However, in this study we do not discriminate between captured ISOs and fragments of ISOs that were tidally disrupted during capture.

Instead of the fitting function~\eqref{eq:f(X):fit} we base all calculations in this study on a spline fit to the numerical $f(X)$ of \citetalias{paper1}.

\section{ISO capture rates}
\label{sec:capture:rate}

\begin{figure}
	\includegraphics[width=\columnwidth]{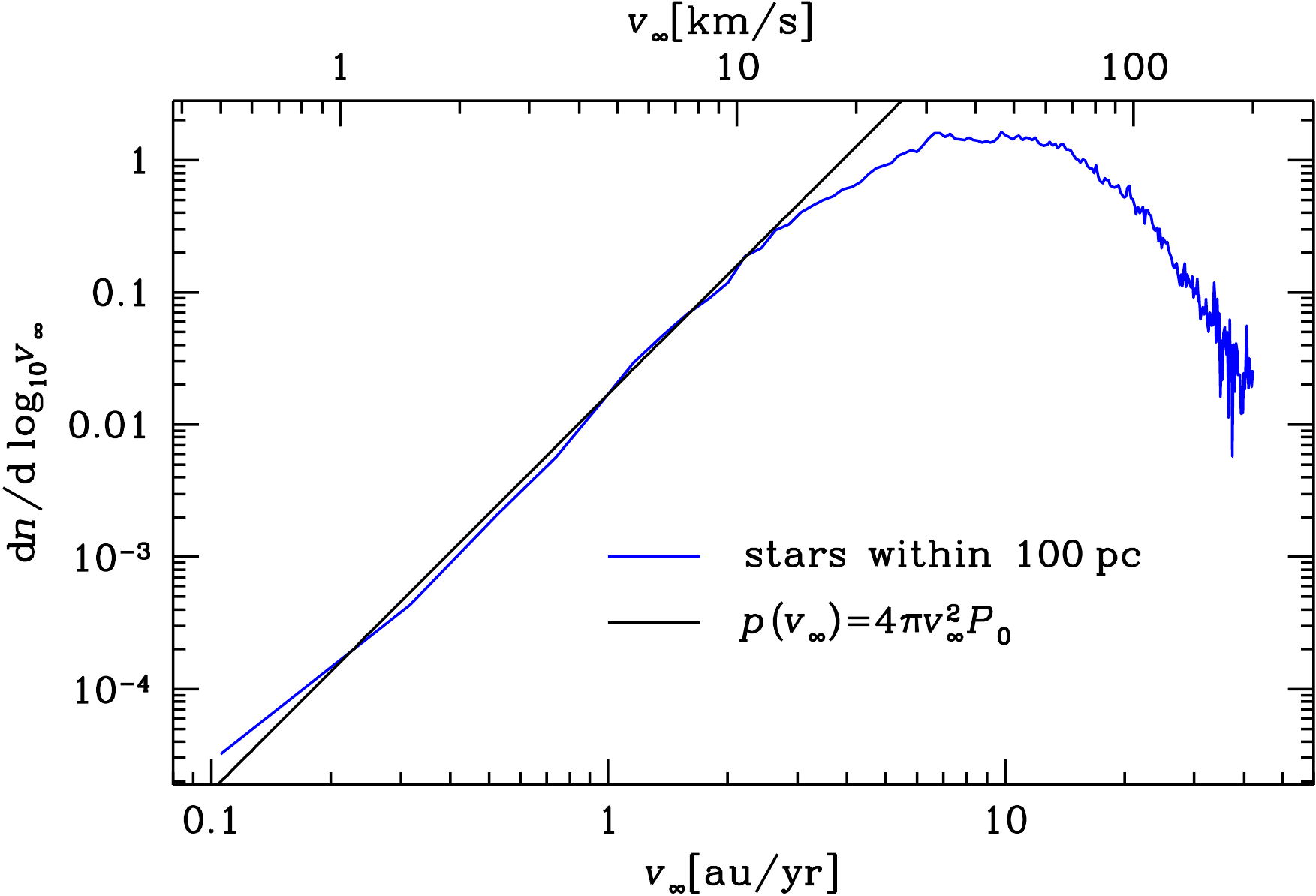}
	\vspace*{-3mm}
	\caption{\label{fig:Gaia}
	Distribution of Stars within 100\,pc from the Sun according to Gaia DR2 in relative speed $v_\infty$ to the Sun. At $v_\infty\lesssim12\,\kms$ this is well described by equation~\eqref{eq:p(v):Taylor}, shown as line.
	}
\end{figure}
\subsection{The cumulative and differential capture rates}
\label{sec:capture:rate:cum+diff}
The rate by which a star-planet binary captures ISOs onto orbits with semi-major axes $\ge a$ is given by
\begin{align}
    \label{eq:Gamma}
    \Gamma(a) = \sub{n}{iso}\,Q(a),
\end{align}
where $\sub{n}{iso}$ is, as before, the number density of ISOs in the Galactic vicinity of the binary (but not enhanced by its gravitational focusing) and
\begin{align}
    \label{eq:Q:rate}
    Q(a) &= \int_0^{\infty} \diff v_\infty\; v_\infty\,p(v_\infty)\;\sigma(v_\infty|\sub{v}a=\sqrt{GM/a})
\end{align}
the \emph{volume capture rate} \citep[or rate function,][]{Heggie1975}. Here, $p(v_\infty)$ is the ISO speed distribution (normalised to unit integral). If $P(\vec{v})$ is their (normalised) space-velocity distribution, then
\begin{align}
    p(v_\infty) &= \int\diff^3\!\vec{v}\,P(\vec{v})\; \delta(v_\infty-|\vec{v}-\vec{v}_0|),
\end{align}
where $\vec{v}_0$ is the space velocity of the capturing binary system. Little is known observationally about the ISO velocity distribution $P(\vec{v})$. Since ISOs form in circum-stellar discs, we surmise that it is very similar to that of the stars\footnote{Unbinding the ISOs from their birth systems via planetary interactions adds $\sim q\sub{v}c^2$ to the velocity dispersion $\sigma_0$ of their parent stars. Since $\sub{v}c\sim\sigma_0$ (tens of $\kms$), this additional heating can be safely neglected. However, the situation may be different if other ejection mechanisms prevail, for example in stellar binaries or close interactions with other stars in the birth cluster.}, which is consistent with the speeds of `Oumouamua and Borisov. Since the capture process will be dominated by ISOs with $v_\infty^2\lesssim q\sub{v}c^2$, much smaller than the velocity dispersion of the stars (and hence by assumption the ISOs), we can Taylor expand $P(\vec{v})$ about $\vec{v}=\vec{v}_0$ to obtain
\begin{align}
    \label{eq:p(v):Taylor}
    p(v_\infty) &= 4\pi P_0^{} v_\infty^2 + O(v_\infty^4),
    &
    P_0^{} &\equiv P(\vec{v}_0).
\end{align}
If we assume $P(\vec{v})$ to be that of the stars in the Solar neighbourhood, then we find from Gaia DR2 $P_0\approx5.5\times10^{-6}\,(\kms)^{-3}\approx5.9\times10^{-4}(\mathrm{au/yr})^{-3}$ and that the approximation~\eqref{eq:p(v):Taylor} holds well for $v_\infty\lesssim12\kms$, see \autoref{fig:Gaia}, which is completely sufficient for the purpose of calculating the capture rates (as we assess more quantitatively below). With this approximation and equation~\eqref{eq:sigma:theory} for the cross-section the volume capture rate~\eqref{eq:Q:rate} becomes
\begin{align}
    \label{eq:Q:a}
    Q(a) &= 2\pi^2 q (GM)^2 P_0\,F\left(\frac{\sub{a}p}{a},q\right),
    \\
    \label{eq:F:z,q}
    F(z,q) &\equiv \int_{z/q}^\infty \diff X\,f(X)\,Y(1,qX-z,z).
\end{align}
At $z\sim1$, corresponding to $a\sim\sub{a}p$, the  factor $Y$ cannot be neglected and implies that $Q\to0$ at $a\to\sub{a}p/2$. But for $z\ll1$, corresponding to $a\gg\sub{a}p$, the integral~\eqref{eq:F:z,q} is dominated by $X<q^{-1}$ and the  function $Y$ unimportant. In this case $Y=1$ gives a good approximation and $F\approx\int_{z/q}^\infty\diff X f(X)$. Hence, for $q\lesssim z\ll 1$, corresponding to $\sub{a}p\ll a \lesssim \sub{a}p/q$, $F\propto z^{-1}\propto a$, while at $z\lesssim q$, corresponding $a\gtrsim \sub{a}p/q$, $F$ depends only weakly on $z$, converging towards $\int_0^\infty\diff X f(X)\approx3.1$. 

\begin{figure}
	\includegraphics[width=\columnwidth]{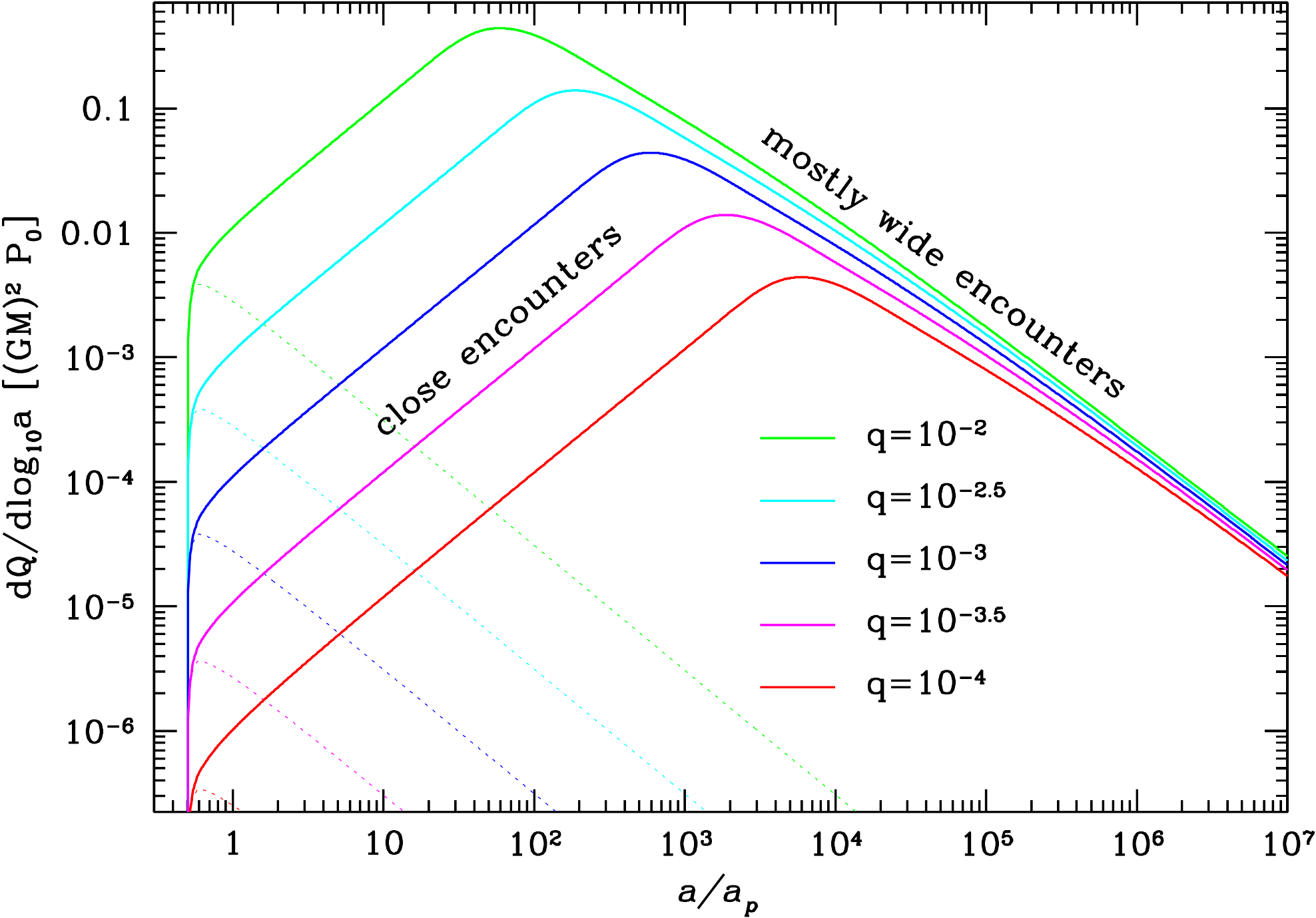}
	\vspace*{-5mm}
	\caption{\label{fig:dQ}
	The differential volume capture rates (equation~\ref{eq:dQ}) for different planet masses. The thin dotted curves are the contributions from the second term in equation~\eqref{eq:dF}. The maximum occurs at $a/\sub{a}p\sim 0.6q^{-1}$, corresponding to the break in $f(X)$ at $X\sim2$. For smaller $a$, captures are due to slingshots during ever closer (and hence less likely) fly-bys. At larger $a$, on the other hand, captures are dominated by wide interactions of ISOs with small incoming speed $v_\infty$, the flux of which diminishes like $v_\infty^3$ as $v_\infty\to0$.
	}
\end{figure}
In the limit $z\ll1$, corresponding to $a\gg\sub{a}p$, the median ISO speed $v_\infty$ captured is obtained at $X=1.4$, while the 90, 95, and 99 percentiles are at $X=8.4$, 16.8, and 80, respectively. For capture by Jupiter, these correspond to $v_\infty=0.5$, 1.2, 1.7, and $3.7\,\kms$, respectively, which agree well with Figure~2 of \cite{HandsDehnen2020}. These numbers are also well below $12\,\kms$ and hence within the range of validity of the Taylor expansion~\eqref{eq:p(v):Taylor}.

We can also obtain the differential volume capture rate onto bound orbits with semi-major axes in the interval $[a,\,a(1+\diff\ln a)]$ as
\begin{align}
    \label{eq:dQ}
    \tdiff{Q}{\ln a} &= 2\pi^2 (GM)^2 P_0^{}\; \frac{\sub{a}p}{a} q \left(-\pdiff{F(z,q)}{z}\right)_{z=\sub{a}p/a}.
\end{align}
From equation~\eqref{eq:F:z,q},
\begin{align}
    \label{eq:dF}
    -\pdiff{F(z,q)}{z}
    &= \frac{1}{q} f\left(\frac{z}{q}\right) Y(1,0,z) \nonumber \\
    &+ \frac{1}{q} \int_0^\infty\diff x\,f\left(X=\frac{x+z}q\right)
    \left[\pdiff{}{x}-\pdiff{}{z}\right]Y(1,x,z).
\end{align}
\autoref{fig:dQ} plots the differential volume capture rate for different planet masses (solid) and separately (dotted) the contributions to $\diff Q/\diff\ln a$ from the second term in equation~\eqref{eq:dF}. Obviously, this term is only important at $a\lesssim\sub{a}p$, i.e.\ for the most bound captures, when collisions (ignored in our treatment) become important.

Our approximation~\eqref{eq:p(v):Taylor} for the ISO speed distribution neglects a similar term. Without that approximation, $-\partial F(z,q)/\partial z$ contains the additional term
\begin{align}
	\label{eq:dF:p}
	\frac{1}{q} \int_0^\infty\diff x\,f\left(X=\frac{x+z}q\right)\,
    \tilde{p}'(x)\, Y(1,x,z),
\end{align}
where the function $\tilde{p}(x)$ is implicitly defined by expressing the ISO speed distribution as $p(v_\infty)=4\pi v_\infty^2 P_0 \tilde{p}(v_\infty^2/\sub{v}c^2)$. In particular, $\tilde{p}\sim1$ and $\tilde{p}'\sim0$ at $x\lesssim1$ for Jupiter or at $x\lesssim2$ for Saturn. Therefore, the integrand in equation~\eqref{eq:dF:p} is always small and the contribution of this term negligible compared to the first term in equation~\eqref{eq:dF}. 

In other words, the differential capture rate is always dominated by captures from small incoming $v_\infty$, regardless of the semi-major axis $a$ of the orbit captured onto, and for $a\gtrsim2\sub{a}p$ an increasingly excellent approximation is
\begin{align}
    \label{eq:dQ:approx}
    \tdiff{Q}{\ln a} &\approx 2\pi^2 (GM)^2 P_0^{}\; \frac{\sub{a}p}{a}
    f\left(\frac{\sub{a}p}{aq}\right).
\end{align}
The minimum semi-major axis onto which an ISO can be captured is half that of the planet, while the most likely (per $\ln a$) is $\sim\sub{a}p/2q$. This corresponds to the maxima in \autoref{fig:dQ} and to the break in $f(X)$ at $X\sim2$: captures onto orbits with $a\lesssim\sub{a}p/q$ are all due to slingshots during close fly-bys with the planet, while those onto orbits with $a\gtrsim\sub{a}p/q$ are dominated by wide interactions. The reduction of the rate (per $\ln a$) at large $a$ is due to the decreasing number of ISOs with small incoming $v_\infty$ (as per equation~\ref{eq:p(v):Taylor}).

\begin{figure}
	\includegraphics[width=\columnwidth]{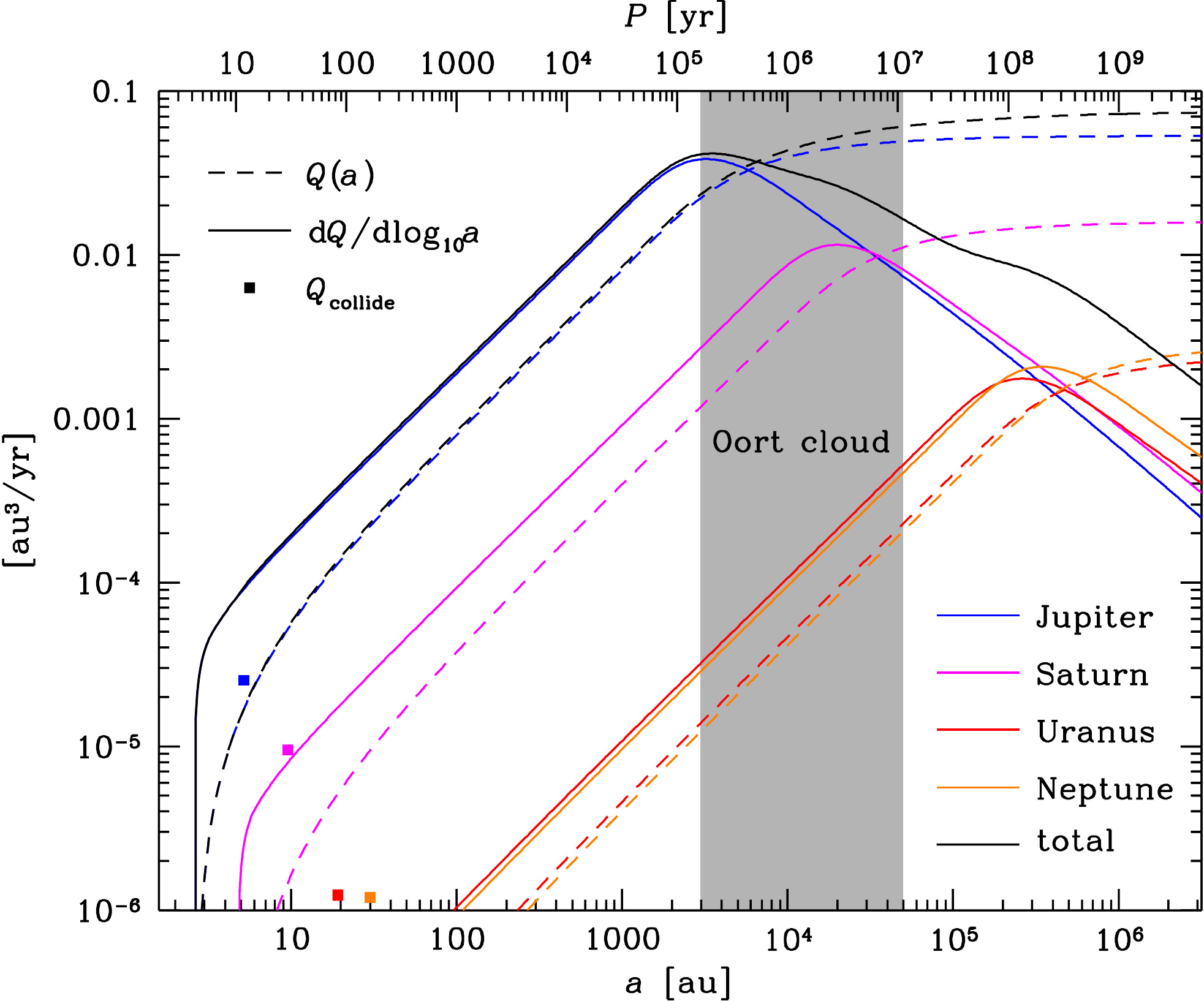}
	\vspace*{-5mm}
	\caption{\label{fig:Q:SS}
	Cumulative (dashed) and differential (solid) volume capture rates of ISOs (assumed to follow the velocity distribution of stars in the Solar neighbourhood) by outer Solar-system planets. The grey region corresponds to the Oort cloud. Also shown are the volume collision rates $Q_{\mathrm{coll}}$ of ISOs with the planets (calculated via equation~\ref{eq:Q:coll}); at $Q(a) < \tfrac12 \sub{Q}{coll}$, the capture rates are overestimated, see also an explanation in the text.
	}
\end{figure}

\subsection{Capture rates by Solar-system planets}
\label{sec:capture:rate:SS}

\autoref{fig:Q:SS} plots the ISO volume capture rates for the Solar-system planets. Obviously, Jupiter dominates the capture into the Solar system at $a\lesssim 10^4\,$au. The maximum for $\diff Q/\diff \ln a$ for Jupiter occurs at $a=2725\,$au or period $P=142263\,$yr, in excellent agreement with results from numerical simulations by \citeauthor{HandsDehnen2020} (\citeyear{HandsDehnen2020}, Fig.~1), and coinciding with the inner edge of the Oort cloud \citep{Oepik1932, Oort1950} at $a=3000\,$au \citep{DuncanQuinnTremaine1987}.
At larger $a$, captures by Saturn also become important and contribute about half of all captures at $a\gtrsim20000\,$au, though the cumulative (i.e.\ total) capture rate onto orbits with $a<50000\,$au is still dominated by Jupiter. In \autoref{fig:Q:SS}, the Oort cloud with limits 3000 and 50000\,au (e.g.\ \citeauthor{DuncanQuinnTremaine1987}) is indicated in grey. Objects with $a>50000\,$au are only very weakly bound to the Sun and likely to be lost within $\sim1$ orbit due to Galactic tides or perturbation by passing stars (or the planets when at perihelion).

We also show in \autoref{fig:Q:SS} (as solid squares) the volume collision rates calculated as
\begin{align}
    \label{eq:Q:coll}
    \sub{Q}{coll} = \int_0^\infty \diff v_\infty\,v_\infty\,p(v_\infty)\, \sub{\sigma}{coll} 
\end{align}
with the cross-section $\sub{\sigma}{coll}$ of equation~(12) of \citetalias{paper1} and the speed distribution $p(v_\infty)$ taken to be that of the stars in the Solar neighbourhood as plotted in \autoref{fig:Gaia}. Since collisions as alternative to captures have been neglected in our derivation of the capture rates, the comparison of the rates $\sub{Q}{coll}$ and $Q(a)$ allows us to assess the limits for the validity of this negligence. Had the planet zero size, the colliding trajectories suffered an energy change, half of which would be negative but possibly insufficient for capture. Hence, we expect less than half of the collisions to be falsely accounted as captures. For Jupiter, for example, that implies that the capture rates are possibly overestimated for $a\lesssim4\,$au, while for Saturn this limit is $\sim20\,$au and $\sim200\,$au for Uranus and Neptune. In all these cases, the errors are insignificant for the total capture rates, but are important for capture onto very tightly bound orbits.

\autoref{tab:Q:SS} summarises the Solar-system volume capture rates into the Oort cloud ($3000\,\mathrm{au}<a<50000\,$au) and into orbits closer than that cloud. As evident from \autoref{fig:Q:SS}, captures into orbits at $a<3000\,$au are due to close encounters with Jupiter (and 5\% also with Saturn), while those into the Oort cloud are dominated by wide encounters with Jupiter. Remarkably, the rate~\eqref{eq:Q:coll} of ISO collisions with the Sun, as computed in \citetalias{paper1}, is 0.1737\,au$^3\,$yr$^{-1}$, about three times the combined capture rate of all planets: passing ISO are three times more likely to fall into the Sun than to be captured into the Solar-system orbits.

\begin{table}
    \caption{Volume capture rates in au$^3$/yr of ISOs by the Solar system planets into orbits with semi-major axes $a<3000\,au$ and $3000\,\mathrm{au} < a < 50000\,$au (Oort cloud).}
    \label{tab:Q:SS}
    \centering
    \begin{tabular}{lll}
        planet & $Q(a<3000\,$au) & 
                 $Q(3000\,\mathrm{au}< a < 50000\,$au) \\
    \hline
        Jupiter & 0.022 & 0.027 \\
        Saturn  & 0.0012 & 0.0099 \\
        Uranus  & 0.000014 & 0.00022 \\
        Neptune & 0.000012 & 0.00019 \\
    \hline
        total   & 0.0234 & 0.0373 \\
    \hline
    \end{tabular}
\end{table}

\subsection{Time dependency of the capture rate}
\label{sec:capture:time}
Combining equation~\eqref{eq:Gamma} with~\eqref{eq:Q:a} or~\eqref{eq:dQ}, we see that the ISO capture rate $\Gamma\propto \sub{n}{iso} P_0$, which is nothing but the ISO phase-space density near the Sun. In other words, if $\sub{F}{iso}(\vec{x},\vec{v})$ is the ISO phase-space density, then $\Gamma\propto\sub{F}{iso\odot}\equiv \sub{F}{iso}(\vec{x}_\odot,\vec{v}_\odot)$. Since ISO dynamics is largely collision-less, it satisfies $\diff\sub{F}{iso}/\diff t=0$ (collision-less Boltzmann equation). This implies that the capture rate does not vary along the Galactic orbit of the Sun, i.e.\ over $\lesssim10^9$ years. Of course, generation of ISOs in regions of ongoing star formation violates the collision-less Boltzmann equation, but since the Sun's orbit is $\sim15\,\kms$ from the local standard of rest at all times, this process does not affect $\sub{F}{iso\odot}$.

This argument rests on the validity of the Taylor expansion~\eqref{eq:p(v):Taylor}, which is the basis for the proportionality $\Gamma\propto \sub{F}{iso\odot}$. As we have shown in \Sect{sec:capture:rate:cum+diff}, 99\% of captures into the Solar system occur at $v_\infty\lesssim3.7\,\kms$. Therefore, $\Gamma\propto \sub{F}{iso\odot}$ holds as long as $\sub{F}{iso}$ remains smooth near the Sun on this scale. Since ISOs are generated within circum-stellar systems, we expect $\sub{F}{iso}$ to be smooth as it is the distribution $\sub{F}{*}$ of their parent stars (which is sufficiently smooth as demonstrated in \autoref{fig:Gaia}) convolved with the distribution of ejection speeds, which are expected to be of the order of a few $\kms$ \citep[e.g.][]{HandsEtAl2019} and near-isotropic owed to subsequent perturbations by GMCs. Strictly, we cannot exclude the existence of a star cluster (i.e.\ small-scale structure of $\sub{F}{*}$ harbouring a cloud of ISOs) close to the orbit of the Sun, such that it will pass through the cluster with less than $\Delta v=4\kms$ relative speed. For this to happen within $T=1\,$Gyr, such a cluster must today be within $\Delta v T=4\,$kpc in mainly azimuthal direction and have avoided detection.

On time scales longer than $\sim10^9$ years the capture rate is affected by the dynamical heating of the ISO population (by interactions with GMCs, spirals arms, etc.\ in the same way as the stellar population), including the drifting in of newly formed ISOs and, possibly, their destruction (nitrogen ice fragments are destroyed by cosmic rays in $\sim\;$4-5\,Gyr, \citealt{Desch21}).

Initially, when the Sun was newborn and still within its birth cluster, the ISO phase-space density, and hence the capture rate, may have been much larger than today. However, it appears unlikely that any ISOs captured at that time are still present in the Solar system today rather than having mostly been ejected again, see also \Sect{sec:pop:phase}.

\section{A population of captive exobodies}
\label{sec:pop}
The capture rates into the Solar system derived above raise the question after the present-day population $\sub{N}{iso}(a)$ of ISOs bound to the Solar-system with semi-major axes $<a$. The gravitational dynamics facilitating capture is fundamentally reversible. Therefore, the reverse process, ejection, is as important as capture and must be accounted for when estimating $\sub{N}{iso}(a)$. In \Sect{sec:pop:time}, we do so by balancing the efficiencies of capture and ejection, while in \Sect{sec:pop:phase} we use the concept of phase-space-volume conservation for the same purpose. Finally, in \Sect{sec:pop:spatial} we derive the spatial density of captured ISOs.

\subsection{Balancing capture and ejection}
\label{sec:pop:time}
Captured ISOs stay on average only for some finite time $\sub{T}{stay}(a)$, depending on their orbit. After a time longer than $\sub{T}{stay}$ has passed, captures and ejections balance at a stable steady-state population of
\begin{align}
    \label{eq:N:iso:prod}
    \tdiff{\sub{N}{iso}}{a}=\sub{n}{iso} \tdiff{Q}{a} \sub{T}{stay}(a)
\end{align}
of captive ISOs with semi-major axis $a$. We now distinguish between ISOs captured onto orbits with semi-major axes $a\lesssim3000\,$au and $a\gtrsim3000\,$au, i.e.\ below or above the maximum of $\diff Q/\diff \ln a$ (see \autoref{fig:Q:SS}).

\subsubsection{Captive population at very long periods}
These latter orbits at $a\gtrsim3000\,$au are similar to those of long-period comets (LPCs) and have perihelia close to the orbit of Jupiter \citep[see also][]{HandsDehnen2020}. At each perihelic passage, they will suffer an energy change comparable in magnitude to that of the wide encounter that bound most of them to the Sun in the first place. Therefore, ISOs captured into such orbits are likely to become unbound within a only a few orbits, though some get scattered onto smaller semi-major axes. We may assume a typical length of stay of $\sub{T}{stay}\sim5$ orbital periods. Then the rate of captive ISOs on LPC orbits to visit the inner Solar system is $\sim5$ times the capture rate, i.e.\ $\sim \sub{n}{iso}0.2\,\mathrm{au^3/yr}$. For `Oumuamua-type ISOs, this means roughly two per century. The total population of these captured ISOs in the Oort cloud counts to 
\begin{align}
    \sub{n}{iso}
    \int_{3000\,\mathrm{au}}^{50000\,\mathrm{au}}
    \sub{T}{stay}(a) \tdiff{Q}{a}\,\diff a \sim 300000 \,\mathrm{au}^3\, \sub{n}{iso},
\end{align}
which for `Oumuamua-type ISOs gives $\sim30000$, negligible compared to $\sim10^{11-12}$ objects in the Oort cloud. 

\subsubsection{Captive population at $a\lesssim3000\,$au}
The situation is quite different for ISOs captured into orbits with semi-major axis $a\lesssim 3000\,$au, because these orbits cannot be unbound by a single wide encounter. Instead, unbinding these orbits requires a much less likely close encounter, comparable to the one that bound them to the Sun them in the first place. The differential capture rate from a single planet for these orbits can be estimated using the strong interaction limit $f(X)=\frac83X^{-2}$:
\begin{align}
    \label{eq:Q:SS:small:a}
    \tdiff{Q}{a} \approx \frac{16\pi^2}{3} q^2 \frac{(GM)^2}{\sub{a}{p}} P_0,
\end{align}
which for Jupiter gives $8.5\times10^{-6}\,\mathrm{au^2/yr}$ at $4\mathrm{au}\lesssim a<2000\,$au and 20 times less for Saturn (at $a\gtrsim20\,$au). At each perihelion passage, these captives will again come within $\sub{a}{p}$ of the Sun and hence may suffer another slingshot. In fact at each passage through the inner Solar system, they will suffer some change (of either sign) in their semi-major axis, resulting in a random-walk. In addition, there is always the chance of being ejected. For simplicity, we ignore the random walk in energy and estimate the length of stay as the orbital period divided by the chance of ejection \citepalias[see Appendix~A of][]{paper1}. If ignoring Saturn, this gives $\sub{T}{stay}\sim 7\,$Myr for Jupiter-crossing orbits at $a=10\,$au and decreases like $a^{-1/2}$ towards larger semi-major axes. Inserting this into equation~\eqref{eq:N:iso:prod} gives the long-term population
\begin{align}
    \label{eq:Niso:rate}
    \tdiff{\sub{N}{iso}}{a} \simeq 190\,\sub{n}{iso}
    \left(\frac{a}{\mathrm{au}}\right)^{-1/2} \mathrm{au}^2.
\end{align}
However, this is an underestimate, since we have ignored Saturn's contribution. Even though its capture rate is only 5\% that of Jupiter, Saturn's ejection rate is also smaller, so that its contribution to the resident population of ISOs is larger than 5\%. Accounting for the influence of Saturn, Uranus, and Neptune on the resident population is  more straightforward using the concept of phase-space volume, which we now pursue.

\subsection{Phase-space based estimates}
\label{sec:pop:phase}
As we have seen already in \Sect{sec:capture:time}, the capture rate is directly proportional to the ISOs phase-space density at the phase-space position of the Sun. The ISOs are simply tracers of the corresponding phase-space volume, some of which is captured into the Solar system. Since according to Liouville's theorem phase-space volume is conserved, the Solar system must also eject phase-space at the same rate as it captures.

The bound phase-space volume of Solar-system orbits with semi-major axis $a$ and eccentricity $e$ is
\begin{align}
    \label{eq:V:ae}
    \diff V = (2\pi)^3 (GM_\odot)^{3/2} a^{1/2}\diff a\,e\,\diff e
\end{align}
\citep[see][Problem 4.8]{BinneyTremaine2008}. Most of the corresponding elliptic orbits are quite stable, meaning that phase-space volume is not, or very rarely, exchanged with other orbits. However, there are two `porous' regions, where phase-space volumes regularly change orbits, mixing phase-space between them. One is at the outer edge of the Solar system at $a\gtrsim50000\,$au, where Galactic tides and passing stars affect the trajectories. The other is the inner Solar system, in particular near Jupiter, where trajectories suffer some change of their energy, implying exchanges of phase-space volume between all such orbits (and resulting in the aforementioned random walk of ISOs).

At both of these regions, phase-space volume is also exchanged with the unbound phase space and these exchanges occur equally in both directions. For example, at $a\gtrsim50000\,$au the same volume is ejected as is captured, only that the ejected volume is populated with Oort-cloud comets, while the captured volume is largely empty.

\subsubsection{The phase-space throughput time}
\label{sec:pop:phase:time}
We first calculate the time scale over which an orbit captures as much as its own volume from unbound phase space:
\begin{align}
    \label{eq:T:thru}
    \sub{T}{thru}(a) = 
    \sub{F}{iso,\odot} \tdiff{V}{a} \bigg/ \tdiff{\Gamma}{a} \approx
    P_0 \tdiff{V}{a} \bigg/ \tdiff{Q}{a}.
\end{align}
In principle, we should calculate $\sub{T}{thru}$ as function of both $a$ and $e$. Instead, in a first estimate we ignore the eccentricity dependence and assume that all orbits crossing a planet orbit, i.e.\ with eccentricity
\begin{align}
    \label{eq:emin}
    e > |1-\sub{a}{p}/a|,
\end{align}
capture ISOs with equal probability at given semi-major axis. We return to the validity of this assumption below.

Integrating equation~\eqref{eq:V:ae} over eccentricities satisfying equation~\eqref{eq:emin}, we obtain the phase-space volume in these orbits for $a>\sub{a}{p}/2$ as
\begin{align}
    \label{eq:V:a}
    \tdiff{V}{a} &=  (2\pi)^3 (GM_\odot)^{3/2} a^{-1/2} \sub{a}{p} \left[1-\frac{\sub{a}{p}}{2a}\right].
\end{align}
Note that the total phase-space volume at given semi-major axis increases like $a^{1/2}$, but the planet-crossing sub-volume decreases like $a^{-1/2}$ at $a\gg\sub{a}p$. The phase-space throughput time for planet-crossing orbits at $a\lesssim2000\,$au then follows from equations~\eqref{eq:Q:SS:small:a} and~\eqref{eq:T:thru} as
\begin{align}
    \label{eqs:T:thru:J,S}
    \sub{T}{thru}(a) &= \frac3{4} \frac{\sub{T}{p}}{q^2}
    \left(\frac{\sub{a}p}{a}\right)^{1/2}
    \left[1-\frac{\sub{a}p}{2a}\right],
\end{align}
independent of $P_0$. Here, $\sub{T}p$ is the orbital period of the planet. This estimate neglects the fact that collisions with the planet reduce the captures at small $a$ as discussed in \Sect{sec:capture:rate:SS} and \autoref{fig:Q:SS}. For Jupiter and Saturn
\begin{subequations}
\begin{align}
    \sub{T}{thru,J}(a) &\approx 2.2\times10^{7\phantom{0}}    \left[1-\frac{\sub{a}{J}}{2a}\right]
    \left(\frac{a}{\mathrm{au}}\right)^{-1/2}\,\mathrm{yr},
    \\
    \sub{T}{thru,S}(a) &\approx 8.3\times10^{8\phantom{0}}       \left[1-\frac{\sub{a}{S}}{2a}\right]
    \left(\frac{a}{\mathrm{au}}\right)^{-1/2}\,\mathrm{yr}.
\end{align}
\end{subequations}
These time scales are remarkably short, even for orbits at small semi-major axes and decrease to 0.5\,Myr and 19\,Myr, respectively, at $a=2000\,$au. For orbits crossing both Jupiter and Saturn, the throughput time is even shorter as slingshots by both planets contribute (one must add the throughput rates $\sub{T}{thru}^{-1}$)\footnote{For Uranus and Neptune, the equivalent calculation gives $\sub{T}{thru} \sim 1.5 \times 10^{11} (\mathrm{au}/a)^{1/2}$ and $\sim 2.5 \times 10^{11} (\mathrm{au}/a)^{1/2}$ years, too long for the corresponding phase-space to be representative of the unbound states, even at $a=2000\,$au and if the rates from Uranus and Neptune can be combined.}.

In reality the capture cross-section $\sigma$ and hence capture rate is not uniform in eccentricity, such that through-put times vary with eccentricity to the same degree as $\sigma$ does. However, even with a factor ten variation, $\sub{T}{thru}\lesssim2\,$Gyr for $a>4\,$au (Jupiter) and $a>20\,$au (Saturn), which is all we need for our argument below.

\begin{figure}
	\includegraphics[width=\columnwidth]{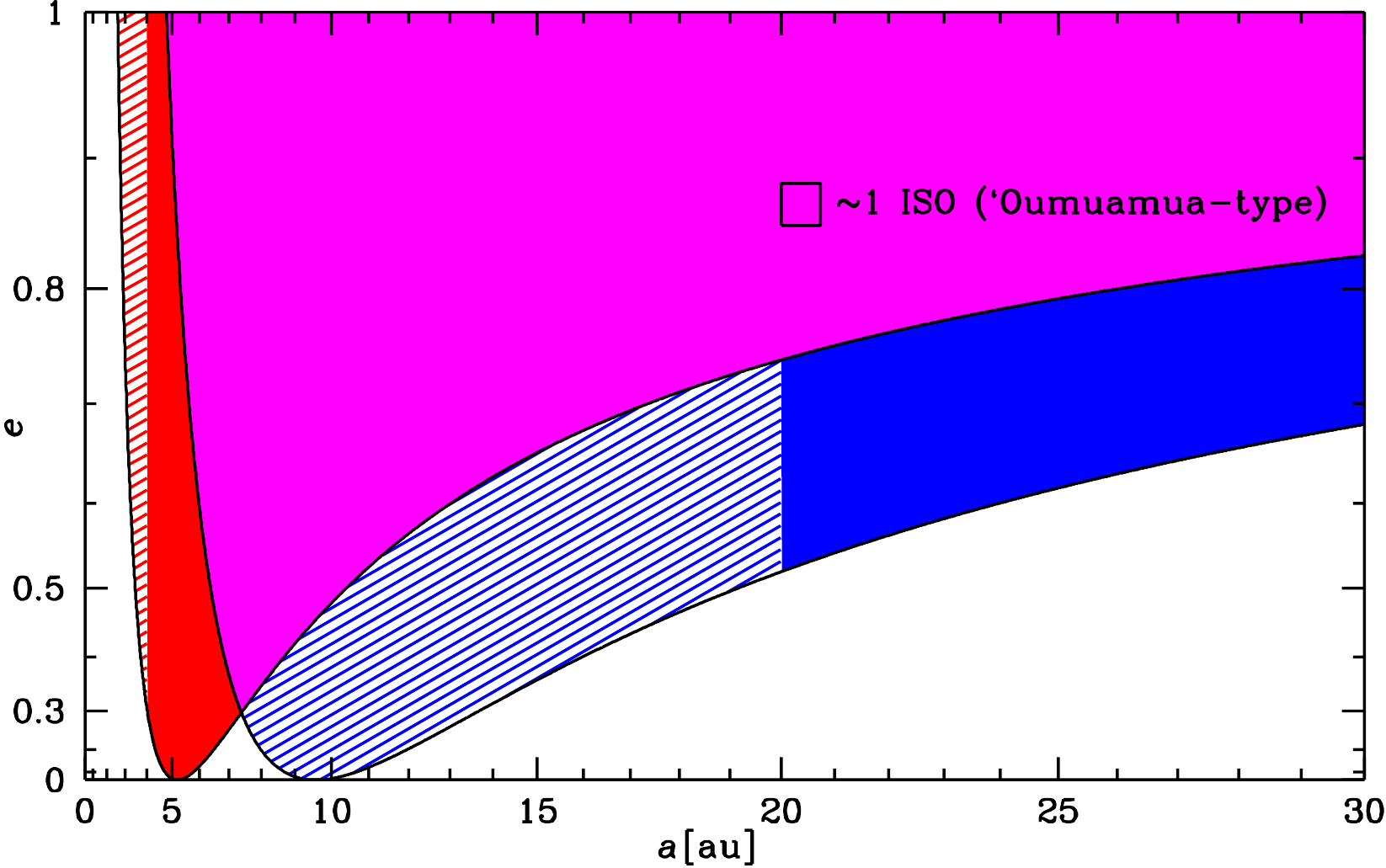}
	\vspace*{-5mm}
	\caption{\label{fig:phase}
    Semi-major axis and eccentricity of orbits crossing Jupiter (red), Saturn (blue), or both (pink). Swing-bys with these planets capture unbound phase-space onto these orbits outside the shaded regions (where most capturing trajectories would penetrate the planet, resulting in collisions instead). The axes are stretched (linear in $a^{3/2}$ and $e^2$) such that equal areas correspond to equal phase-space volumes. The square gives the volume expected to be occupied by one ISO on average.
	}
\end{figure}
\subsubsection{The ISO population}
\label{sec:pop:phase:pop}
On time scales longer than $\sub{T}{thru}$, the capture of phase-space volume from unbound phase-space in the immediate vicinity of the Sun leads to a complete mixing of the corresponding phase-space elements. Therefore, phase-space on orbits where $\sub{T}{thru}\lesssim2\,$Gyr will contain the same density of ISOs as the unbound phase space in the Solar vicinity, namely $\sub{F}{iso,\odot}=\sub{n}{iso} P_0$. From our estimates above, this holds for all orbits at $a\gtrsim4\,$au which cross Jupiter and all orbits at $a\gtrsim20\,$au that cross Saturn. 

\autoref{fig:phase} shows the corresponding $(a,e)$ space in a representation that maps equal area to equal phase-space volume. The number of resident ISOs in the corresponding phase-space then follows from the respective volume~\eqref{eq:V:a} as
\begin{align}
    \label{eq:Niso:phase}
    \tdiff{\sub{N}{iso}}{a}
    &\approx 190 \sub{n}{iso}
    \left[1-\frac{\sub{a}{J}}{2a}\right] \left(\frac{a}{\mathrm{au}}\right)^{-1/2}\,\mathrm{au}^2
    &&\text{for $4\,\mathrm{au}\lesssim a\lesssim20\,$au}, \\ \nonumber
    &\approx 350 \sub{n}{iso}
    \left[1-\frac{\sub{a}{S}}{2a}\right] \left(\frac{a}{\mathrm{au}}\right)^{-1/2}\,\mathrm{au}^2
    &&\text{for $20\,\mathrm{au}\lesssim a\lesssim2000\,$au}.
\end{align}
This result for Jupiter alone is identical to our first estimate~\eqref{eq:Niso:rate} based on balancing capture and ejection rates. Technically, this is because the throughput and remain times, $\sub{T}{thru}$ and $\sub{T}{stay}$, are identical, but conceptually these two population counts are calculated in very different ways, with the latter method arguably being cleaner and less approximate.

For $\sub{n}{iso}=0.1\,\mathrm{au}^{-3}$, the square in \autoref{fig:phase} corresponds to the phase-space volume that contains on average one ISO. Thus several such ISOs are expected to reside within bound orbits at $a<10\,$au at any time.

\begin{figure}
    \begin{center}
	    \includegraphics[width=\columnwidth]{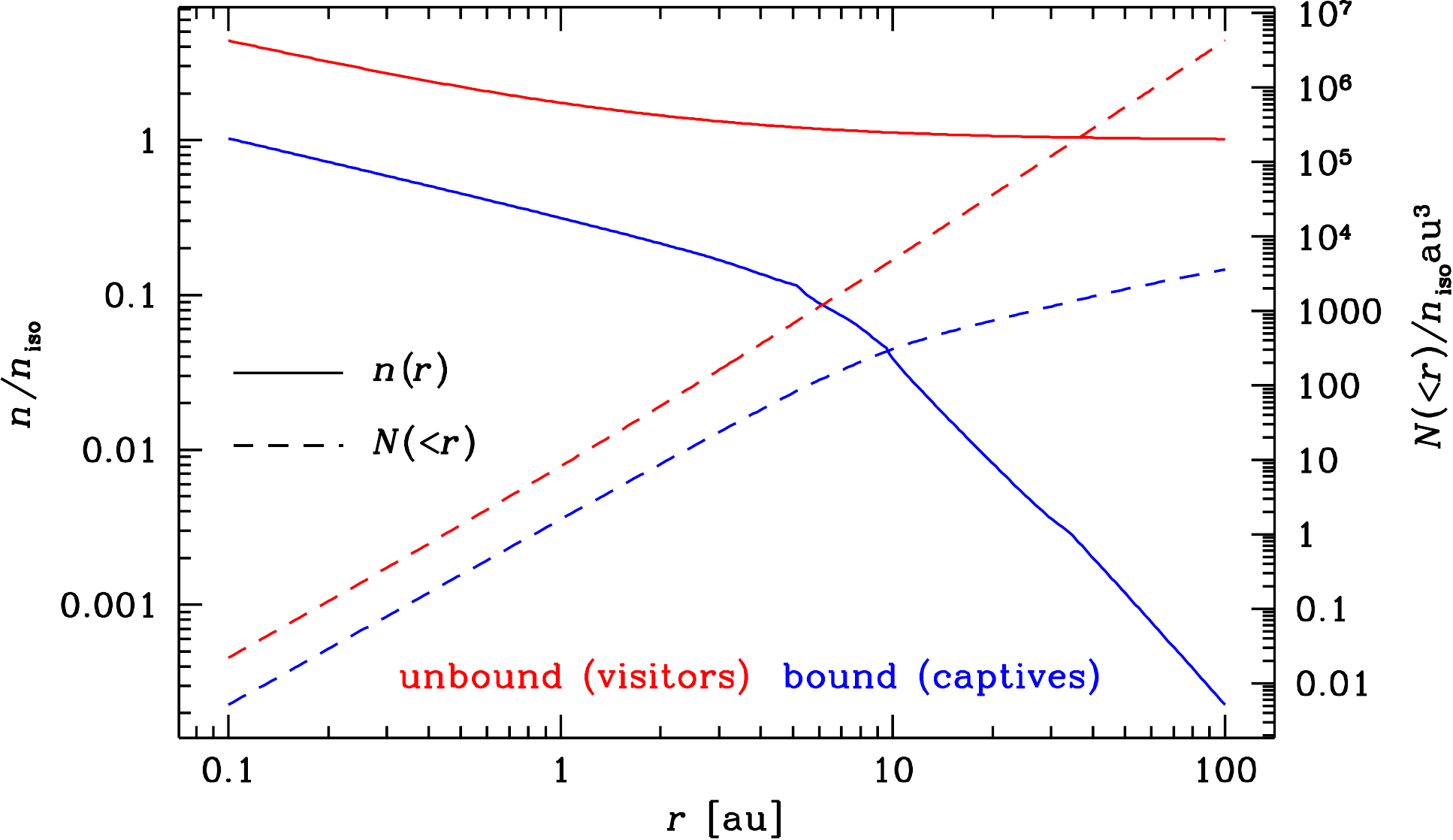}
    \end{center}
	\vspace*{-3mm}
	\caption{\label{fig:dens}
	Space density $n(r)$ and cumulative number $N({<}r)$ of ISOs visiting the Solar system on hyperbolic orbits (red, equation~\ref{eq:n:unbound}) and of ISOs captured onto elliptic orbits (blue, equation~\ref{eq:n:bound}), assuming that Jupiter crossing orbits with $4\,\au<a<2000\,\au$ and Saturn crossing orbits with $20\,\au<a<2000\,\au$ (see \autoref{fig:phase}) contain the same phase-space density as the ISOs in the Galactic vicinity of the Sun and that the ISO velocities are distributed as for the stars.
	}
\end{figure}
\subsection{ISO spatial density}
\label{sec:pop:spatial}
The above estimates provide the number of ISOs on bound orbits as function of their semi-major axis $a$ and eccentricity $e$. We now translate this into the number density at given radius $r$ from the Sun. The distribution function of ISOs resident in the Solar system is, according to our model, constant at $f(a,e)=\sub{F}{iso,\odot}$ for 
\begin{align}
    \label{eq:DF:elimit}
    e^2 > 
    \begin{cases}
    (1-\sub{a}{J}/a)^2
    & \text{for $a_0 < a < a_1$}, \\
    (1-\sub{a}{S}/a)^2
    & \text{for $a_1 < a < a_2$}
    \end{cases}
\end{align}
and zero elsewhere, where $a_0=4\,\au$, $a_1=20\,\au$, and $a_2=2000\,\au$ (this corresponds to the solid coloured regions in \autoref{fig:phase}). The number density of ISOs bound to the Solar system then follows by integrating over all velocities after some algebra as
\begin{align}
    \label{eq:n:bound}
    n(r) = & \frac{4\pi}3(GM_\odot)^{3/2}\sub{n}{iso}P_0
    \left[ \left(\frac2r-\frac1{a_2}\right)_+^{\frac32}
        -   \left(\frac2r-\frac1{a_0}\right)_+^{\frac32}
        -
    \right. \\ \nonumber & - \left.
            \bigg(\frac{\sub{a}{J}^2}{r^2}-1\bigg)_+^{\frac12}
            \left(\frac1{a_0}-\frac2{\sub{a}{J}+r}\right)_+^{\frac32}
        -   \bigg(1-\frac{\sub{a}{J}^2}{r^2}\bigg)_+^{\frac12}
            \left(\frac2{\sub{a}{J}+r}-\frac1{a_1}\right)_+^{\frac32}
        +
    \right. \\ \nonumber & + \left.
            \bigg(1-\frac{\sub{a}{S}^2}{r^2}\bigg)_+^{\frac12}
            \left(\frac2{\sub{a}{S}+r}-\frac1{a_1}\right)_+^{\frac32}
        -   \bigg(1-\frac{\sub{a}{S}^2}{r^2}\bigg)_+^{\frac12}
            \left(\frac2{\sub{a}{S}+r}-\frac1{a_2}\right)_+^{\frac32}
    \right],
\end{align}
where $(\cdot)_+\equiv\max\{0,\cdot\}$. The number density of visiting ISOs that merely pass through the Solar system (like `Oumuamua and Borisov) on unbound orbits is enhanced over $\sub{n}{iso}$ by gravitational focusing:
\begin{align}
    \label{eq:n:unbound}
    \sub{n}{unbound}(r) &= \sub{n}{iso} \int_0^\infty \left(1+\frac{2GM_\odot}{rv_\infty^2}\right)^{1/2} p(v_\infty)\,\diff v_\infty.
\end{align}
In \autoref{fig:dens}, we plot both densities (relative to $\sub{n}{iso}$) as function of radius. The number of visiting ISOs passing through is everywhere much larger than that of captive ISOs. The density ratio approaches $\sim4$ at radii $\lesssim1\,\au$. The increase $n\propto r^{-1/2}$ at small radii is due to highly eccentric planet-crossing orbits.

Also plotted in \autoref{fig:dens} are the cumulative numbers of ISOs (dashed). For $\sub{n}{iso}=0.1\,\mathrm{au}^{-3}$, we expected one unbound visitor within 1\,au at any time, but only 0.2 captives. At 5\,au, these numbers rise to $\sim60$ and $\sim8$.

\section{Discussion and Conclusions}
\label{sec:conclude}

We calculated the rate of capture of interstellar objects (ISOs) into the Solar system, using the corresponding cross-section as obtained in \citetalias{paper1}. The rate of captures is completely dominated by capturing ISOs with low incoming asymptotic speed $v_\infty$ and hence proportional to the phase-space density $\sub{F}{iso}$ of ISOs in the vicinity the Sun. If $\sub{F}{iso}$ is smooth near the Sun on scales $\sim4\,\kms$, the capture rate does not vary along the Solar orbit through the Milky Way: the enhanced ISO density when crossing the Galactic mid-plane is compensated by a dilution of velocities space around the Sun (increase of velocity dispersion). In particular, the capture rate is not enhanced during passage of the Solar system through a cloud of ISO's \citep[in contrast to statements of previous studies e.g.][]{ClubeNapier1984}, except if it moves with less $\lesssim4\,\kms$ with respect to the Sun. However, such clouds cannot emerge from the ISM (as by-product of star formation or otherwise), since the Solar orbit is always $\sim15\,\kms$ away from the local standard of rest, from which the ISM hardly deviates even in spiral arms.

While little is known about the structure of $\sub{F}{iso}$, simulations of the evaporation of Oort clouds suggest that ISOs form tidal streams around each star \citep{CorreaOttoCalandra2019, PortegiesZwart2021}, in contrast to our assumption of a smooth distribution. However, these studies neglected dynamical heating by GMCs, which dissolves such streams once they reach a length of a few pc (the typical size of GMCs) after only $\sim10^{\text{7}}\,$yr (for drift velocities of  $\lesssim1\,\kms$). Moreover, the relative importance of Oort-cloud evaporation as opposed to ejection by close encounters in the stellar birth cluster \citep[e.g.][]{HandsEtAl2019} remains unclear.

Assuming that the velocity distribution of ISOs follows that of the stars, we calculate the capture rates for the outer Solar system planets. These are dominated by Jupiter, which in 1000 years captures $\sim2$ ISOs onto orbits at $a<3000\,$au if the ISO number density $\sub{n}{iso}\sim0.1\,$au$^{-3}$, while Saturn achieves 5\% of that (for comparison, one ISO falls into the Sun on average every 60 years). However, Saturn still contributes significantly to the population of captive ISOs. This is because its phase-space capture rate is sufficient to replenish the phase-space volume of all Saturn crossing bound orbits within $\sim2\,$Gyr or less. The same is true for Jupiter, so that all bound Solar-system orbits crossing those of Jupiter or Saturn contain ISOs at the same phase-space density as the phase-space in the Solar vicinity. For $\sub{n}{iso}\sim0.1\,$au$^{-3}$, we estimate that there are $\sim8$ captured ISOs within 5\,au at any time, which is small compared to $\sim60$ unbound ISOs in the same volume.

In terms of phase-space volume, ejection from and capture onto a particular Solar-system orbit exactly balance according to Liouville's theorem. This holds for the exchange of phase-space between incoming hyperbolic and planet-crossing elliptic orbits as well as for the exchange of these latter with more stable Solar-system orbits. The total phase-space volumes of these more stable and the planet-crossing orbits are roughly comparable at $a\lesssim2000\,$au, but the exchange rate between them is, by definition of orbital stability, quite small. Therefore, orbital pockets of stability where captured ISOs could remain for an extensive period are highly unlikely to contain ISOs captured by the current Solar-system configuration.

On the other hand, changes to the orbital configuration can alter the probability of ejection and stabilise an ISO after capture. For example, if Jupiter were to migrate inwards after capturing an ISO, the chance of orbit crossing and hence ejection would be reduced. Similarly, at aphelia of $\gtrsim10^4\,$au ISOs may gain angular momentum from passing stars, such that they no longer cross Jupiter's orbit \citep{Oepik1932, Oort1950}.

However, there appears to be no possibility to trap an ISO into an orbit at $a\lesssim2000\,$au for very long: there are no orbital traps, the only traps are collisions, especially with the Sun, when dissipation renders the dynamics irreversible and invalidates Liouville's theorem.

\section*{Acknowledgements}
We thank Scott Tremaine for useful discussions and the reviewer, Simon Portegies Zwart, for helpful suggestions. RS acknowledges generous support by a Royal Society University Research Fellowship.

\section*{Data Availability}
No data were generated for this study.

\bibliographystyle{mnras} \bibliography{capture}

\label{lastpage}
\end{document}